\documentclass
[aps,prl,twocolumn,showpacs,lengthcheck,preprintnumbers]{revtex4}%
\usepackage{amsfonts}
\usepackage{amsmath}
\usepackage{amssymb}
\usepackage{graphicx}%
\usepackage{longtable}
\providecommand{\U}[1]{\protect\rule{.1in}{.1in}}
\begin{document}
\preprint{Phys. Rev. E \textbf{88}, 042916 (2013)}
\title{Effective equations for matter-wave gap solitons in higher-order transversal states}
\author{A. Mu\~{n}oz Mateo}
\email{ammateo@ull.es}
\author{V. Delgado}
\email{vdelgado@ull.es}
\affiliation{Departamento de F\'{\i}sica Fundamental II, Facultad de F\'{\i}sica,
Universidad de La Laguna, E-38206 La Laguna, Tenerife, Spain}
\date{8 June 2013}


\begin{abstract}
We demonstrate that an important class of nonlinear stationary solutions of
the three-dimensional (3D) Gross-Pitaevskii equation (GPE) exhibiting 
nontrivial transversal configurations can be found and characterized in terms 
of an effective one-dimensional (1D) model. 
Using a variational approach we derive effective equations of lower
dimensionality for BECs in $(m,n_{r})$ transversal states (states featuring a
central vortex of charge $m$ as well as $n_{r}$ concentric zero-density rings
at every $z$ plane) which provides us with a good approximate solution of the
original 3D problem. Since the specifics of the transversal dynamics can be
absorbed in the renormalization of a couple of parameters, the functional form
of the equations obtained is universal. The model proposed finds its principal
application in the study of the existence and classification of 3D gap
solitons supported by 1D optical lattices, where in addition to providing a
good estimate for the 3D wave functions it is able to make very good
predictions for the $\mu(N)$ curves characterizing the different fundamental
families. We have corroborated the validity of our model by comparing its
predictions with those from the exact numerical solution of the full 3D GPE.

\end{abstract}
\maketitle

%\preprint{ }

%\author{A. Mu\~{n}oz Mateo and V. Delgado}

\section{I. INTRODUCTION}

Matter-wave solitons are an important class of nonlinear excitations in
Bose-Einstein condensates (BECs) \cite{Carre2,RevDS}. They are coherent,
spatially localized solutions of the Gross-Pitaevskii equation (GPE) resulting
from a balance between linear dispersion and nonlinearity. Moreover, they are
the atomic analogs of the optical solitons of nonlinear optics, with which
they share many similarities \cite{KivGOP,Kiv2D,RMP-Malomed}. This analogy has
proved to be very fruitful and has stimulated much theoretical and
experimental work in the field of Bose-Einstein condensation. A particular
example of the interplay between the two fields which have attracted
considerable attention in recent years is that of matter-wave gap solitons
\cite{Kon1,Mor1}. These solitons, first observed experimentally in Ref.
\cite{Ober1}, are localized nonlinear structures which can be realized in BECs
loaded in one-dimensional (1D) optical lattices, and are characterized by a 
chemical potential lying in the forbidden band gaps of the underlying lineal 
problem.

Most theoretical treatments of matter-wave gap solitons in 1D optical lattices
have been carried out in a 1D setting \cite{GS1D,Kiv1,Abd1,Mal1,Wu1-2}. This
approximation is justified when the condensate is subject to a transversal
confinement so strong that its radial dynamics is frozen to the zero-point
quantum fluctuations of the corresponding ground state. Under such
circumstances, the axial dynamics can be accurately described in terms of the
1D GPE. However, as the transversal confinement decreases or the condensate
density increases, the contribution from higher-order radial modes becomes
more and more important so that transversal excitations can no longer be
neglected \cite{VDB7}. This situation occurs frequently in practice.

In general, the study and characterization of matter-wave dark or gap solitons
proceeds in different and well separated stages \cite{Carre1}. The first step
consists in establishing the existence of such nonlinear solutions which, for
solitons with a nontrivial radial configuration, requires solving the
nonlinear eigenvalue problem associated with the three-dimensional (3D) GPE. 
This is a complex problem which is commonly addressed numerically by using a 
Newton continuation method in terms of the chemical potential of the condensate. 
In a subsequent step one usually is interested in the stability properties of 
the nonlinear states previously found. Such properties can be determined from 
a linear stability analysis based on the numerical solution of the 
Bogoliubov--de Gennes equations or by monitoring the long-time behavior of the 
condensate after a sudden small perturbation, which in turn requires the 
numerical integration of the equations of motion. In a final stage, one may be 
interested in the stabilization of certain unstable solutions, for instance, 
by changing the system configuration, the number of particles, or by 
introducing convenient pinning potentials or even a second component with 
repulsive interspecies interactions. In general, the different stages of the 
analysis have to be tackled separately by using different specific computational 
techniques. In what follows we will focus on the first stage. More specifically, 
we will be interested in the search for axially localized nonlinear stationary 
solutions of the 3D GPE exhibiting a higher-order radial configuration. While, 
in principle, this is a 3D problem where both axial and radial degrees of 
freedom can play an equally relevant role \cite{GS-3D}, we will show in this 
work that, rather unexpectedly, it can be addressed in terms of an equivalent
effective problem of lower dimensionality. Such dimensionality reduction is
particularly interesting because the computational complexity of a Newton
continuation method increases rapidly as one passes from 1D to 3D.

Effective equations of reduced dimensionality have proven to be a valuable
tool in the description of BECs in highly anisotropic geometries. This class
of systems, which can be realized by applying a strong confinement potential in
one or two spatial directions, has received considerable attention in recent
years \cite{Olsha1,Kett1,Das1,Strin1,VDB1,Guerin1,Isa1}. In particular, they
are especially relevant for applications of trapped Bose condensates in
matter-wave interferometry \cite{Shin2,Schumm1,Wang1} or in generation and
manipulation of matter-wave dark \cite{Burger1,Brand1} and gap solitons
\cite{Ober1}. Since highly anisotropic potentials induce two very different
time scales in the condensate dynamics, in these systems the computational
effort of a fully 3D treatment can become prohibitively high \cite{PRL06,Mott}%
. This is so due to the fact that in order to guarantee the convergence of the
numerical procedure one has to accurately resolve the fast degrees of freedom
even though very often only the slow ones are relevant. Fortunately, the
dynamics of dilute quantum gases tightly confined in the radial or the axial
direction becomes effectively one dimensional or two dimensional,
respectively, which allows for a description of the condensate dynamics in
terms of effective equations of lower dimensionality. A number of such 
low-dimensional equations have been proposed
\cite{Vic1,Jack1,Chio1,Gora,Salas1,Kam1,Clark1,Yuka1,Ef1DEqs,Nico1}. Especially
relevant because of their simplicity and practical usefulness in realistic
situations are those derived in Refs. \cite{Salas1} and \cite{Ef1DEqs}. It has
been shown that they can also be obtained from a variational approach based,
respectively, on the appropriate energy and chemical-potential functionals
\cite{Ef1DEqs}. Several variants and extensions of the above two latter
equations has also been proposed
\cite{Modug1,Salas2,Salas3,Salas4,Adhik1,Adhik2,Adhik3,Carre3,Salas5,Adhik4}.
For highly elongated condensates, it has been shown that, within their range
of applicability, the effective equations derived in Refs. \cite{Salas1} and
\cite{Ef1DEqs} can accurately reproduce the experimental results
\cite{Carre3,Kev1,Wel1,Kev2}. In fact, unlike what happens with the standard
1D GPE (which is a nonlinear Schr\"{o}dinger equation with a cubic
nonlinearity), the above mentioned effective equations can account properly
for the contribution of transversal excitations on the axial dynamics of the
condensate through a nonpolynomial nonlinear term. For this reason, they are
particularly well suited for modeling trapped condensates in the dimensional
crossover regime between 1D and 3D, a regime which is especially relevant in
the study of matter-wave nonlinear structures such as dark and gap solitons.

Interestingly, the effective equations derived in this work are a natural
generalization of the effective equations of Refs. \cite{Salas1} and
\cite{Ef1DEqs} to the case where the condensate exhibits a nontrivial
higher-order transversal configuration. While stationary ringlike transversal
excitations have been studied previously (both theoretically and
experimentally) \cite{Kiv-RDS,LCarr-RDS,Kev-RDS1,San-RDS,Kev-RDS2,Liu-RDS} in
BECs axially confined by a harmonic trapping potential (where the axial
degrees of freedom play no relevant role), here we are mainly concerned with a
quite different situation, where, in addition to a nontrivial radial
configuration, the condensates feature a localized self-trapped nonlinear
structure along the axial direction (such as occurs, for instance, in the 
case of matter-wave gap solitons in the weak-radial-confinement regime
\cite{GS-3D}). In such circumstances, the axial degrees of freedom play as
important role as the radial ones and thus must be explicitly incorporated in
the physical description. As we will demonstrate, the effective equations
derived in this work permit one to solve such an essentially 3D problem in 
terms of an equivalent 1D problem whose computational cost is similar to 
that corresponding to solving the standard 1D GPE.

\section{II. EFFECTIVE 1D MODEL}

In the mean-field regime, the physics of dilute Bose-Einstein condensates at
zero temperature, in a confining potential $V(\mathbf{r})$, is completely
characterized by a macroscopic wave function $\psi(\mathbf{r},t)$ that
satisfies the Gross-Pitaevskii equation \cite{GPE}
\begin{equation}
i\hbar\frac{\partial\psi}{\partial t}=\left(  -\frac{\hbar^{2}}{2M}\nabla
^{2}+V(\mathbf{r})+gN\left\vert \psi\right\vert ^{2}\right)  \psi, \label{R-7}%
\end{equation}
where $N$ is the number of atoms, and $g=4\pi\hbar^{2}a/M$ is the interaction
strength determined by the \textsl{s}-wave scattering length $a$.

Equation (\ref{R-7}) has been proven to accurately reproduce the dynamics of
mean-field BECs under realistic experimental conditions \cite{PRL06,Mott}. Our
aim here is to look for a simpler effective equation of lower dimensionality
which, under certain conditions, permit us to find good approximate solutions
of the equation above. To this end, we start by assuming the adiabatic
approximation to be applicable. More specifically, we shall assume that the
axial degrees of freedom evolve so slowly in time in comparison with the
transverse degrees of freedom that, at every instant $t$, the latter ones can
relax almost instantaneously to a stationary state compatible with the axial
configuration. Under these circumstances, correlations between axial and
radial motions are negligible and the wave function can be factorized in the
form \cite{Jack1}
\begin{equation}
\psi(\mathbf{r},t)=\varphi(\mathbf{r}_{\bot};n_{1})\phi(z,t), \label{R-8}%
\end{equation}
where $\mathbf{r}_{\bot}=(x,y)$ and $n_{1}$ is the axial linear density,%
\begin{equation}
n_{1}(z,t)\equiv N\int d^{2}\mathbf{r}_{\bot}|\psi(\mathbf{r}_{\bot}%
,z,t)|^{2}=N|\phi(z,t)|^{2}. \label{R-9}%
\end{equation}
In the above equations, both $\psi$ and $\varphi$ have been normalized to unity.

Assuming, as is commonly the case, a separable confining potential
$V(\mathbf{r})=V_{\bot}(\mathbf{r}_{\bot})+V_{z}(z)$, after substituting Eq.
(\ref{R-8}) into the GPE and averaging over the fast degrees of freedom one
finally arrives at the following effective 1D equation governing the slow
axial dynamics of the condensate \cite{Ef1DEqs}:%
\begin{equation}
i\hbar\frac{\partial\phi}{\partial t}=-\frac{\hbar^{2}}{2M}\frac{\partial
^{2}\phi}{\partial z^{2}}+V_{z}(z)\phi+\mu_{\bot}(n_{1})\phi, \label{I-4}%
\end{equation}
where $\mu_{\bot}(n_{1})$ is the local chemical potential satisfying the
stationary transverse GPE%
\begin{equation}
\left(  -\frac{\hbar^{2}}{2M}\nabla_{\bot}^{2}+V_{\bot}(\mathbf{r}_{\bot
})+gn_{1}\left\vert \varphi\right\vert ^{2}\right)  \varphi=\mu_{\bot}%
(n_{1})\varphi. \label{I-5}%
\end{equation}
In what follows, we shall restrict ourselves to condensates confined in
cylindrically symmetric traps which are harmonic in the radial direction,%
\begin{equation}
V(\mathbf{r})=\frac{1}{2}M\omega_{\bot}^{2}r_{\bot}^{2}+V_{z}(z). \label{I-6}%
\end{equation}
Under these circumstances, in the ideal-gas linear regime ($an_{1}%
\rightarrow0$), the stationary transverse equation (\ref{I-5}) can be exactly
solved in terms of generalized Laguerre polynomials, $L_{n_{r}}^{(|m|)}%
(\rho^{2})$, and its solutions take the form%
\begin{equation}
\varphi_{n_{r}}^{(m)}(\rho,\theta)=N_{n_{r}}^{(m)}e^{im\theta}\rho
^{|m|}e^{-\rho^{2}/2}L_{n_{r}}^{(|m|)}(\rho^{2}), \label{S-1}%
\end{equation}
where $N_{n_{r}}^{(m)}$\ is the normalization constant,%
\begin{equation}
N_{n_{r}}^{(m)}\equiv\sqrt{\frac{n_{r}!}{\pi a_{\bot}^{2}(n_{r}+|m|)!}}\;;
\label{S-2}%
\end{equation}
$\theta$ is the azimuthal angle; and $\rho\equiv r_{\bot}/a_{\bot}$ is the
dimensionless radial coordinate, with $a_{\bot}=\sqrt{\hbar/M\omega_{\bot}}$
being the harmonic-oscillator length. In the equation above, $n_{r}%
=0,1,2,\ldots$ is the radial quantum number, and $m=0,\pm1,\pm2,\ldots$ is the
axial angular momentum quantum number, which accounts for the presence of an
axisymmetric vortex of charge $m$ (with $m=0$ corresponding to the absence of
vortices). Using the following explicit expression for the generalized
Laguerre polynomials \cite{Abramo}:%
\begin{equation}
L_{n_{r}}^{(m)}(x)=\sum_{k=0}^{n_{r}}(-1)^{k}%
\begin{pmatrix}
n_{r}+m\\
n_{r}-k
\end{pmatrix}
\frac{x^{k}}{k!}, \label{S-3}%
\end{equation}
Eq. (\ref{S-1}) can be rewritten in the form%
\begin{equation}
\varphi_{n_{r}}^{(m)}(\rho,\theta)\!=\!N_{n_{r}}^{(m)}e^{im\theta}e^{-\rho
^{2}/2}\sum_{k=0}^{n_{r}}\!\frac{(-1)^{k}}{k!}\!%
\begin{pmatrix}
n_{r}+|m|\\
n_{r}-k
\end{pmatrix}
\!\rho^{2k+|m|} \label{S-4}%
\end{equation}
These wave functions, which are stationary states of the underlying linear
problem with energies%
\begin{equation}
E=(2n_{r}+|m|+1)\hbar\omega_{\bot}, \label{S-4b}%
\end{equation}
will be our starting point in the search for solutions of Eq. (\ref{I-5})
in the nonlinear regime.

In principle, solving Eq. (\ref{I-5}) is equivalent to finding the stationary
points of the energy functional%
\begin{equation}
\frac{E[\varphi]}{N}\equiv\!\!\int\!d^{2}\mathbf{r}_{\bot}\left(
\!\frac{\hbar^{2}}{2M}\left\vert \nabla_{\bot}\varphi\right\vert ^{2}+V_{\bot
}\left\vert \varphi\right\vert ^{2}+\frac{1}{2}gn_{1}\left\vert \varphi
\right\vert ^{4}\!\right)  \!, \label{S-5}%
\end{equation}
with $n_{1}=N/L$ being the condensate linear density. However, as was
demonstrated in Ref. \cite{Ef1DEqs}, for repulsive interatomic interactions
($a>0$), when the search for the stationary points of $E[\varphi]$ is
restricted to a subspace of trial functions (as is usually the case) a
variational approach based on the chemical-potential functional%
\begin{equation}
\mu_{\bot}[\varphi]\!\equiv\!\!\int\!d^{2}\mathbf{r}_{\bot}\varphi^{\ast
}\!\left(  \!-\frac{\hbar^{2}}{2M}\nabla_{\bot}^{2}+V_{\bot}(\mathbf{r}_{\bot
})+gn_{1}\left\vert \varphi\right\vert ^{2}\!\right)  \!\varphi, \label{I-9}%
\end{equation}
can produce more simple and yet more accurate results than the usual
variational approach based on the energy functional. For this reason, in what
follows we will look for variational solutions of Eq. (\ref{I-5}) in terms of 
both functionals.

It is natural to choose as trial functions those given in Eq. (\ref{S-4}) with
the substitution $a_{\bot}\rightarrow\Gamma a_{\bot}$, where $\Gamma$ is a
variational parameter characterizing the condensate width. Solutions of this
kind are analytic continuations of the linear eigenfunctions that bifurcate
from the different linear energy levels as one enters the nonlinear regime.
Thus, associated with every set of quantum numbers $(m,n_{r})$ there exists a
one-parameter family of nonlinear stationary solutions that reduce to the
stationary states of the underlying linear problem, Eq. (\ref{S-4}), in the
limit $an_{1}\rightarrow0$. This enables us to categorize the solutions of Eq.
(\ref{I-5}) into families, and associate a given family with a pair
$(m,n_{r})$. Note that the existence of such solutions implies the
conservation in the nonlinear regime of the topology that the condensates
exhibit in the ideal-gas linear regime. In fact, in passing from linear to
nonlinear stationary states of a given family not only is the vortex charge
$m$ conserved (which is a direct consequence of the underlying rotational
symmetry), but so is the number ($n_{r}$) of concentric zero-density rings
(radial nodes). And this occurs despite the fact that the nonlinear solutions
are, in general, superpositions of many linear modes with different $n_{r}$
quantum numbers.

%%%%  Table 1  %%%%

%\begin{table*}[tb]
\begin{table*}[t]
\caption{Parameters $\alpha$, $\eta$, and $\gamma$ for the first few radial
modes $(|m|,n_{r})$ as follow from Eqs. (\ref{S-7})--(\ref{S-10}).}%
\label{Tabla1}%
\centering\vspace{0.2cm}
\begin{ruledtabular}
\begin{tabular}
%[c]{cccccccccc}\hline\hline
[c]{cccccccccc}
$(|m|,n_{r}):$ & (0,0) & (1,0) & (2,0) & (0,1) & (1,1) & (2,1) & (0,2) &
(1,2) & (2,2)\\\hline
$\alpha:$ & 1 & 2 & 3 & 3 & 4 & 5 & 5 & 6 & 7\\
$\eta:$ & 1 & 1/2 & 3/8 & 1/2 & 5/16 & 1/4 & 11/32 & 15/64 & 199/1024\\
$\gamma:$ & 1 & 1/4 & 1/8 & 1/6 & 5/64 & 1/20 & 11/160 & 15/384 &
199/7168\\%
%\hline\hline
\end{tabular}
\end{ruledtabular}
\end{table*}
%\end{table*}

Substituting the trial functions into the energy functional (\ref{S-5}), after
a rather cumbersome calculation, one obtains%
\begin{equation}
\frac{E(n_{1};\Gamma)}{N}=\frac{\hbar\omega_{\bot}}{2\Gamma^{2}}\left(
\alpha+\alpha\Gamma^{4}+2\eta an_{1}\right)  , \label{S-6}%
\end{equation}
where the parameters $\alpha$ and $\eta$, which contain all the dependence on
the quantum numbers $(|m|,n_{r})$, are given by%
\begin{align}
\alpha &  \equiv\sum\limits_{i,j=0}^{n_{r}}C_{i}C_{j}%
\,(1+i+j+|m|)!\;,\label{S-7}\\
\eta &  \equiv\!\sum\limits_{i,j,k,l=0}^{n_{r}}\!C_{i}C_{j}C_{k}%
C_{l}\,2^{-(i+j+k+l+2|m|)}(i\!+\!j\!+\!k\!+\!l\!+\!2|m|)!\;, \label{S-8}%
\end{align}
with%
\begin{equation}
C_{i}=\frac{(-1)^{i}\sqrt{n_{r}!(n_{r}+|m|)!}}{i!(n_{r}-i)!(i+|m|)!}.
\label{S-9}%
\end{equation}
At this point, it is most convenient to introduce a new parameter $\gamma$,
defined as
\begin{equation}
\gamma\equiv\eta/\alpha, \label{S-10}%
\end{equation}
and to rewrite Eq. (\ref{S-6}) in terms of $\alpha$ and $\gamma$. Minimization
of this latter equation with respect to the variational parameter then leads
to the following $z$-dependent condensate width:%
\begin{equation}
\Gamma=\left(  1+2\gamma an_{1}\right)  ^{1/4}. \label{S-11}%
\end{equation}
Substituting back in Eq. (\ref{S-6}) and using that $\mu_{\bot}(n_{1}%
)=\partial E/\partial N$, one finds the following transverse local chemical
potential:%
\begin{equation}
\mu_{\bot}(n_{1})=\hbar\omega_{\bot}\alpha\frac{1+3\gamma an_{1}}%
{\sqrt{1+2\gamma an_{1}}}. \label{S-12}%
\end{equation}
After inserting the above expression into the axial equation (\ref{I-4}), one
finally arrives at%
\begin{equation}
i\hbar\frac{\partial\phi}{\partial t}=-\frac{\hbar^{2}}{2M}\frac{\partial
^{2}\phi}{\partial z^{2}}+V_{z}(z)\phi+\hbar\omega_{\bot}\alpha\frac{1+3\gamma
aN\left\vert \phi\right\vert ^{2}}{\sqrt{1+2\gamma aN\left\vert \phi
\right\vert ^{2}}}\phi. \label{S-13}%
\end{equation}
This is an effective 1D equation that governs the axial dynamics of elongated
BECs whose transversal state is a nonlinear mode belonging to the $(m,n_{r})$
family. Such transversal states, which exhibit an axisymmetric vortex of
charge $m$ and $n_{r}$ concentric zero-density rings, at every instant $t$ and
$z$ plane are indistinguishable from a (stationary) axially homogeneous
condensate satisfying Eq. (\ref{I-5}) with a linear density $n_{1}%
(z,t)=N|\phi(z,t)|^{2}$ and quantum numbers $(m,n_{r})$.

Equation (\ref{S-13}) is generally applicable to condensates with both
attractive ($a<0$) and repulsive ($a>0$) interatomic interactions. In the
latter case, as already said, a somewhat simpler equation can be derived by
using directly the chemical-potential functional instead of the usual energy
functional. Indeed, after substituting the above trial functions into Eq.
(\ref{I-9}), one obtains%
\begin{equation}
\mu_{\bot}(n_{1};\Gamma)=\frac{\hbar\omega_{\bot}}{2\Gamma^{2}}\alpha\left(
1+\Gamma^{4}+4\gamma an_{1}\right)  , \label{S-14}%
\end{equation}
where the parameters $\alpha$ and $\gamma$ are those defined previously in
Eqs. (\ref{S-7})--(\ref{S-10}). Minimizing $\mu_{\bot}$ with respect to the
variational parameter $\Gamma$, one now finds the $z$-dependent condensate
width%
\begin{equation}
\Gamma=\left(  1+4\gamma an_{1}\right)  ^{1/4}, \label{S-15}%
\end{equation}
and, after substituting back in Eq. (\ref{S-14}), one obtains the following
local chemical potential:%
\begin{equation}
\mu_{\bot}(n_{1})=\hbar\omega_{\bot}\alpha\sqrt{1+4\gamma an_{1}}.
\label{S-16}%
\end{equation}
Inserting this expression into Eq. (\ref{I-4}), we find%
\begin{equation}
i\hbar\frac{\partial\phi}{\partial t}=-\frac{\hbar^{2}}{2M}\frac{\partial
^{2}\phi}{\partial z^{2}}+V_{z}(z)\phi+\hbar\omega_{\bot}\alpha\sqrt{1+4\gamma
aN\left\vert \phi\right\vert ^{2}}\phi. \label{S-17}%
\end{equation}
This equation, like Eq. (\ref{S-13}), is an effective 1D equation governing
the axial dynamics of elongated BECs with $a>0$ in a $(m,n_{r})$ transversal 
state. Equations (\ref{S-13}) and (\ref{S-17}), as well as their
time-independent counterparts, are the central results of this work. It is
interesting to note that these effective equations exhibit the same functional
form irrespective of the $(m,n_{r})$ transversal state involved. In fact, the
details of the transversal dynamics enter the above equations only through the
numerical values of the parameters $\alpha$ and $\gamma$, which account for
the dilution effect that the different transversal configurations induce in
the radially averaged condensate density. The numerical values of $\alpha$ and
$\gamma$ for the first few $(|m|,n_{r})$ pairs, obtained from Eqs.
(\ref{S-7})--(\ref{S-10}), are given in Table \ref{Tabla1}. Note, in
particular, that the parameter $\alpha$ can be conveniently rewritten as%
\begin{equation}
\alpha=(2n_{r}+|m|+1). \label{S-18}%
\end{equation}
This formula is a direct consequence of the fact that the local chemical
potential (\ref{S-16}) must reduce to the (linear) energy (\ref{S-4b}) in the
$an_{1}\rightarrow0$ limit.

In the particular case $n_{r}=0$, which corresponds to condensates whose
transversal state contains only an axisymmetric vortex of charge $m$,
expressions (\ref{S-7}) and (\ref{S-8}) become%
\begin{equation}
\alpha=(1+|m|), \label{S-19}%
\end{equation}%
\begin{equation}
\eta=(1+|m|)\gamma=\frac{(2|m|)!}{2^{2|m|}(|m|!)^{2}}\equiv\beta_{m}^{-1},
\label{S-20}%
\end{equation}
so that the equations of motion (\ref{S-13}) and (\ref{S-17}) reduce,
respectively, to the effective equations first derived in Refs. \cite{Salas3}
and \cite{Ef1DEqs}. It is thus clear that the equations derived in this work
are natural generalizations of the effective equations obtained previously and
reduce to them in the proper limit.

Simple closed expressions for the parameter $\gamma$ can also be obtained in
other particular cases. For instance, setting $n_{r}=1$ in Eq. (\ref{S-8})
yields%
\begin{equation}
\gamma=\eta/\alpha=\frac{(2+3|m|)}{4(1+|m|)(3+|m|)}\beta_{m}^{-1}\;;
\label{S-21}%
\end{equation}
while, for $n_{r}=2$, after a rather lengthy calculation, one finds%
\begin{equation}
\gamma=\frac{44+(95+41|m|)|m|}{64(1+|m|)(2+|m|)(5+|m|)}\beta_{m}^{-1}.
\label{S-22}%
\end{equation}

The validity of the effective equations derived above relies on both the
applicability of the adiabatic approximation and the robustness of the
$(m,n_{r})$ transversal state involved. For general dynamical problems with
$n_{r}>0$, such conditions may represent a rather severe limitation. There
are, however, physical situations where one is primarily concerned with the
existence and classification of stationary nonlinear excited states. This
occurs, for instance, in the search for localized nonlinear structures such as
matter-wave dark and gap solitons. While the study of this kind of system has
commonly been performed in a 1D setting in which the radial state plays no
relevant role, one may also be interested in nonlinear excited modes
corresponding to higher-order transversal states. In principle, such analysis
would require solving numerically the stationary 3D GPE, which is a
computationally expensive task. Moreover, the numerical solution of the
corresponding nonlinear eigenvalue problem---based, in most cases, on a
continuation method---depends critically on the possibility of starting the
iterative procedure with a sufficiently good initial estimation for
\emph{both} the axial and the radial parts of the wave function, which are not
known \textit{a priori}. As we shall see, the time-independent counterparts of 
the effective 1D equations (\ref{S-13}) and (\ref{S-17}) prove to be valuable
tools that permit one to overcome the above difficulties. Since for a 
stationary problem the radial degrees of freedom can always follow 
adiabatically the (static) axial configuration, it is clear that, in such 
cases, the validity of the adiabatic approximation is guaranteed. In fact, the 
stationary versions of the above effective 1D equations have a broader range 
of applicability than initially expected, as demonstrated by the fact that 
they are even able to reproduce rather accurately the physical properties of 
disk-shaped gap solitons in higher-order transversal states.

\section{III. MATTER-WAVE SOLITONS IN HIGHER-ORDER TRANSVERSAL STATES}

While small differences between results from the effective equations
(\ref{S-13}) and (\ref{S-17}) do exist (see Ref. \cite{Ef1DEqs}), in practice,
however, they are not very important, so that, from a quantitative point of
view, the choice of which one to use is a matter of personal preference. For
condensates with repulsive interatomic interactions ($a>0$) it may be
advantageous to use Eq. (\ref{S-17}) because it is somewhat simpler and,
furthermore, permits one to obtain fully analytical expressions for a number 
of important stationary physical magnitudes \cite{Ef1DEqs}. For $a<0$,
however, one necessarily has to resort to Eq. (\ref{S-13}) which has the
additional advantage that it is also valid for $a>0$. In what follows we shall
consider condensates with $a>0$ and shall restrict ourselves to Eq.
(\ref{S-17}). To demonstrate its utility in the search for matter-wave dark
and gap solitons in a 3D regime, next we shall compare, for a number
of\ representative examples, the predictions from this equation with those
from the exact numerical solution of the full 3D GPE with no approximations.
To this end, throughout this work we consider a zero-temperature $^{87}$Rb
condensate (\textsl{s}-wave scattering length $a=5.29$ nm) in a harmonic
radial trap, subject to different axial potentials $V_{z}(z)$, and look for
stationary solutions of Eq. (\ref{S-17}) of the form%
\begin{equation}
\phi(z,t)=\phi_{0}(z)\exp(-i\mu t/\hbar),\label{III-1}%
\end{equation}
where $\mu$ is the condensate chemical potential and its corresponding density
amplitude $\phi_{0}(z)$ satisfies the stationary equation%
\begin{equation}
\left(  \!\!-\frac{\hbar^{2}}{2M}\frac{\partial^{2}}{\partial z^{2}}%
\!+\!V_{z}(z)\!+\!\hbar\omega_{\bot}\alpha\sqrt{1+4\gamma aN\left\vert
\phi_{0}\right\vert ^{2}}\right)  \!\phi_{0}\!=\!\mu\phi_{0}.\label{III-2}%
\end{equation}
Starting from a convenient initial guess and using $\mu$ as a continuation
parameter, excited solutions of a nonlinear differential equation such as this
one can be found numerically by means of a Newton continuation method. As we
shall see below, by solving the above effective 1D problem one can find good
approximate solutions of the corresponding 3D problem. Indeed, not only does
the solution of Eq. (\ref{III-2}) accurately provide the chemical potential
$\mu$ of the 3D condensate as a function of its particle content $N$ (a
functional dependence that is of primary interest in the characterization and
classification of matter-wave gap solitons into different families), but also
gives, via Eqs. (\ref{R-8}) and (\ref{III-1}), an estimate for the
corresponding 3D wave function:%
\begin{equation}
\psi_{n_{r}}^{(m)}(\mathbf{r},t)=\varphi_{n_{r}}^{(m)}(\mathbf{r}_{\bot}%
;n_{1})\phi_{0}(z)\exp(-i\mu t/\hbar),\label{III-3}%
\end{equation}
where $\varphi_{n_{r}}^{(m)}$ is given by Eqs. (\ref{S-1}) and (\ref{S-2})
with the substitution $a_{\bot}\rightarrow\Gamma a_{\bot}$ and, as follows
from Eqs. (\ref{R-9}) and (\ref{S-15}), the $z$-dependent dimensionless
condensate width $\Gamma$ takes the form%
\begin{equation}
\Gamma=\left(  1+4\gamma aN\left\vert \phi_{0}(z)\right\vert ^{2}\right)
^{1/4}.\label{III-4}%
\end{equation}
Stationary solutions of the full 3D GPE (\ref{R-7}) have been obtained
numerically in a similar manner. As already said, however, in this case it is
essential to start from an initial guess which is sufficiently good in both
the axial and the radial directions. In this regard, a good strategy is to use
the above 1D estimate (for a relatively small value of $\mu$), Eq.
(\ref{III-3}), as the starting point for the 3D problem. We have implemented
the numerical continuation method in terms of a Laguerre-Fourier spectral
basis and have carefully checked the convergence and accuracy of our results
by using different basis sets.

%%%%%%%%%%%%%%%%%%%%%%%%%%%%%%%%%%%%%%%%%%%%%%%%%%%%%%%%%%%%%%%%%%%%%%%%%%%%%%%

\begin{figure}[ptb]
\begin{center}
\includegraphics[
width=8.4cm
]{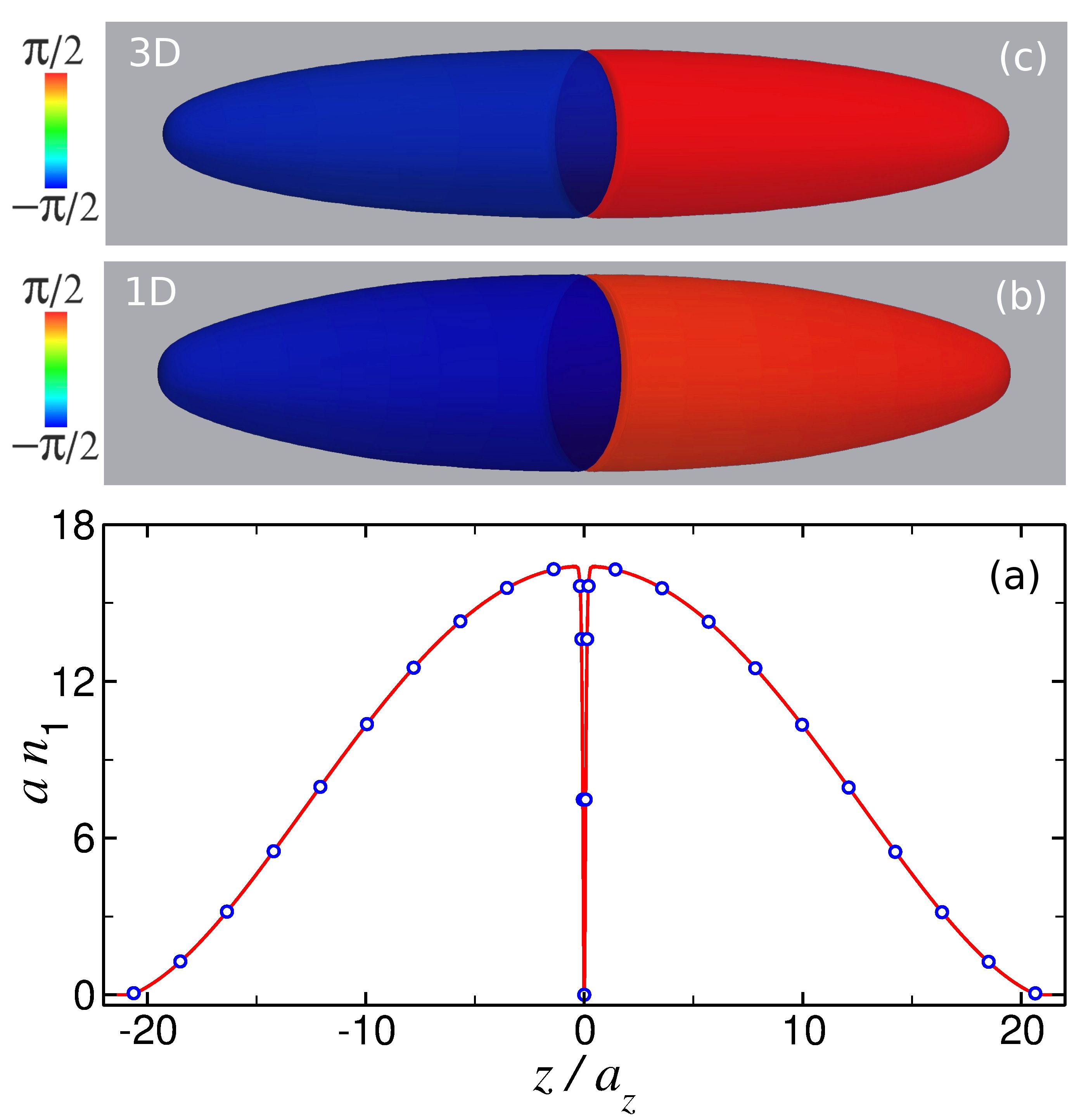}
\end{center}
\caption{
(Color online) 3D morphology of a stationary axial dark soliton in a $(0,0)$
transversal state generated in a $^{87}$Rb condensate with $2\times10^{5}$
atoms confined in a harmonic trap with frequencies $\omega _{z}=2\pi\times15$
Hz and $\omega_{\bot}=2\pi\times450$ Hz. (a) Dimensionless axial density $%
an_{1}$ as follows from the effective 1D equation (\ref{III-2}) (solid red
line) along with the corresponding 3D results (open circles) obtained from
the full 3D GPE. The two upper panels show phase-colored density isosurfaces
(taken at $5\%$ of the respective maxima) representing the 3D wave functions
as obtained from (b) the effective 1D model and (c) the full 3D GPE. Panels
(b) and (c) have been radially magnified by a factor of $5$. The color maps
in these panels show that in both cases the condensate phase takes the
constant value $-\pi /2$ to the left of the central density notch and $+\pi
/2$ to the right. Lengths are given in terms of the axial oscillator length $%
a_{z}=$ $2.78$ $\mu$m.}%
\label{Fig1}%
\end{figure}

%%%%%%%%%%%%%%%%%%%%%%%%%%%%%%%%%%%%%%%%%%%%%%%%%%%%%%%%%%%%%%%%%%%%%%%%%%%%%%%

\begin{figure}[ptb]
\begin{center}
\includegraphics[
width=8.4cm
]{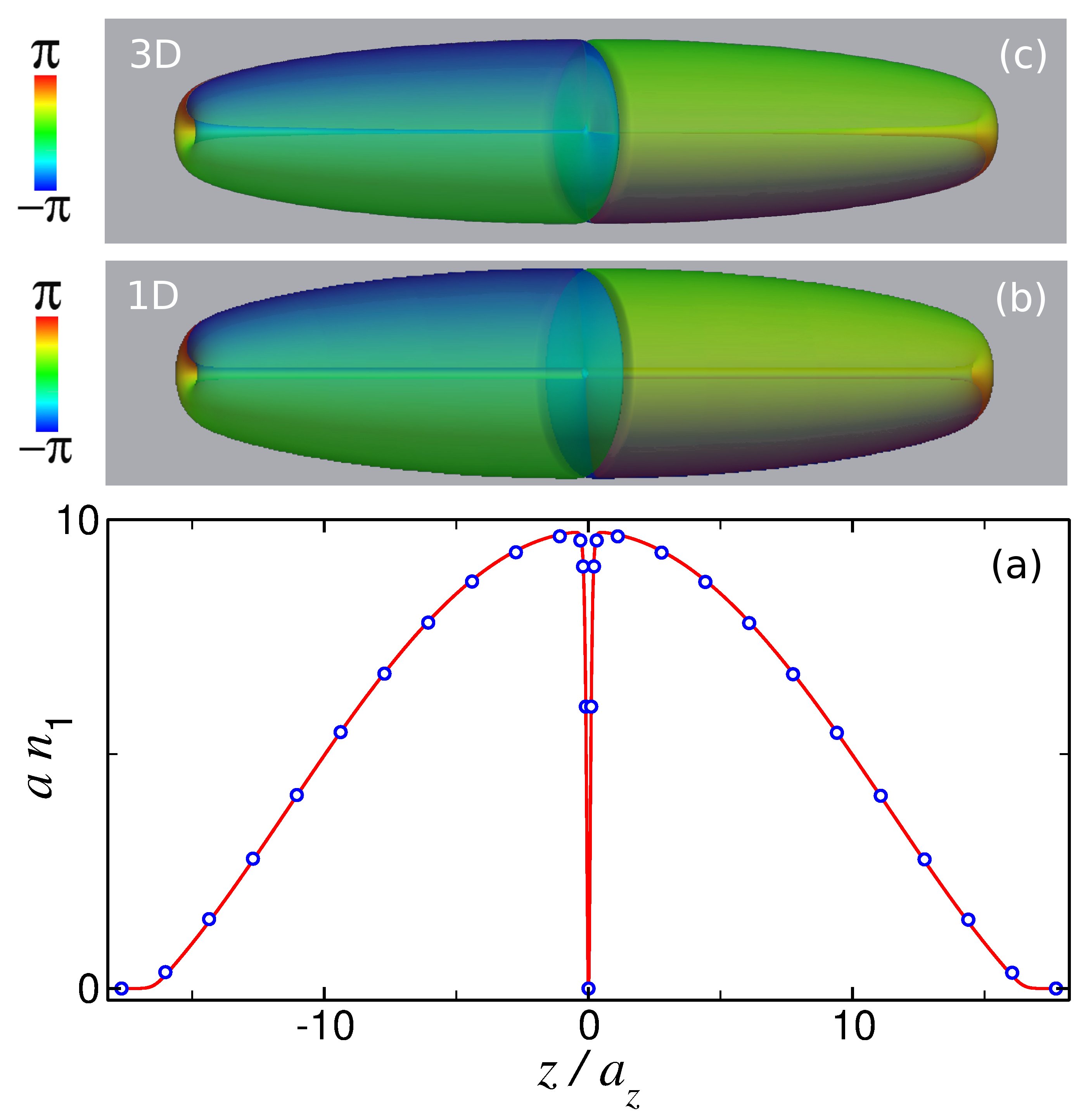}
\end{center}
\caption{
(Color online) Same as Fig. 1 but for a stationary axial dark soliton in a 
$(1,0)$ transversal state containing $10^{5}$ atoms. The color maps in 
panels (b) and (c) show that in both cases the condensate phase varies 
continuously from $-\pi $ to $+\pi $ around the vortex axis, while along 
the axial direction, for each azimuthal angle it exhibits a $\pi $-phase slip 
across the notch of zero density.}%
\label{Fig2}%
\end{figure}

%%%%%%%%%%%%%%%%%%%%%%%%%%%%%%%%%%%%%%%%%%%%%%%%%%%%%%%%%%%%%%%%%%%%%%%%%%%%%%%

Figure \ref{Fig1} shows a stationary axial dark soliton in a $(0,0)$
transversal state generated in a $^{87}$Rb condensate containing
$2\times10^{5}$ atoms confined in a harmonic trap with axial and radial
frequencies $\omega_{z}=2\pi\times15$ Hz and $\omega_{\bot}=2\pi\times450$ Hz,
respectively. Lengths in this figure are given in terms of the corresponding
axial oscillator length $a_{z}=\sqrt{\hbar/M\omega_{z}}=$ $2.78$ $\mu$m.
Figure \ref{Fig1}(a) depicts the dimensionless axial density $an_{1}%
=aN|\phi_{0}(z)|^{2}$ as follows from the effective 1D equation (\ref{III-2})
(solid red line) along with the corresponding 3D results (open circles)
obtained from the full 3D GPE as prescribed by Eq. (\ref{R-9}). Substituting
the density amplitude $\phi_{0}(z)$ obtained from the solution of Eq.
(\ref{III-2}) into Eqs. (\ref{III-3}) and (\ref{III-4}), we have generated 
an effective 3D wave function which is shown in Fig. \ref{Fig1}(b) as an
isosurface of the atom density (taken at $5\%$ of its maximum) where the value
of the phase of the wave function at each point is represented as a color map.
Figure \ref{Fig1}(c) displays the 3D wave function obtained from the numerical
solution of the stationary version of the full 3D GPE (\ref{R-7}) with no
approximations. According to this latter equation, the condensate chemical
potential takes the value $\mu=8.19\;\hbar\omega_{\bot}$, while the effective
1D equation (\ref{III-2}) predicts $\mu=8.16\;\hbar\omega_{\bot}$ (which
implies an error of only $0.35\%$). Even though in this case the transversal
state is topologically trivial, it is nevertheless a linear superposition of
\emph{many} harmonic oscillator modes as reflects the fact that the chemical
potential $\mu$ is much greater than the radial quantum $\hbar\omega_{\bot}$.
In these circumstances, the standard 1D GPE is not applicable. In fact, this
condensate clearly lies in the 3D regime, far away from the quasi-1D regime
which would require $an_{1}\ll1$, a condition that is not met as is evident
from Fig. \ref{Fig1}(a). Along the axial direction, however, both the density
and the phase profiles exhibit the peculiar nontrivial structure of a dark
soliton. Indeed, from Figs. \ref{Fig1}(b) or \ref{Fig1}(c) (which have been 
magnified along the radial direction by a factor of $5$ to facilitate the 
visualization) the characteristic $\pi$-phase slip across the notch of zero 
density is evident.

For the same confinement potential, Fig. \ref{Fig2} shows an axial
dark soliton in a $(1,0)$ transversal state containing $10^{5}$ atoms. This
corresponds to a stationary state of the GPE featuring a vortex of unit charge
in the transversal direction in addition to a dark soliton in the axial
direction. Figure \ref{Fig2}(a) compares the prediction for the axial density
$an_{1}$ (solid red line) obtained from Eq. (\ref{III-2}) (which yields
$\mu=6.56\;\hbar\omega_{\bot}$) with the exact numerical result (open circles)
obtained from the 3D GPE (which yields $\mu=6.61\;\hbar\omega_{\bot}$). The
phase-colored density isosurfaces shown in Figs. \ref{Fig2}(b) and 
\ref{Fig2}(c) represent, respectively, the corresponding 3D wave functions as 
obtained from the effective 1D model and the full 3D GPE. From these figures 
one can appreciate the complex phase structure of this nonlinear stationary 
state, which exhibits an azimuthal $2\pi$-phase slip in every $z$ plane 
besides a $\pi$-phase slip along the axial direction \emph{for each value} of 
the azimuthal angle $\theta$.

%%%%%%%%%%%%%%%%%%%%%%%%%%%%%%%%%%%%%%%%%%%%%%%%%%%%%%%%%%%%%%%%%%%%%%%%%%%%%%%

\begin{figure}[ptb]
\begin{center}
\includegraphics[
width=8.4cm
]{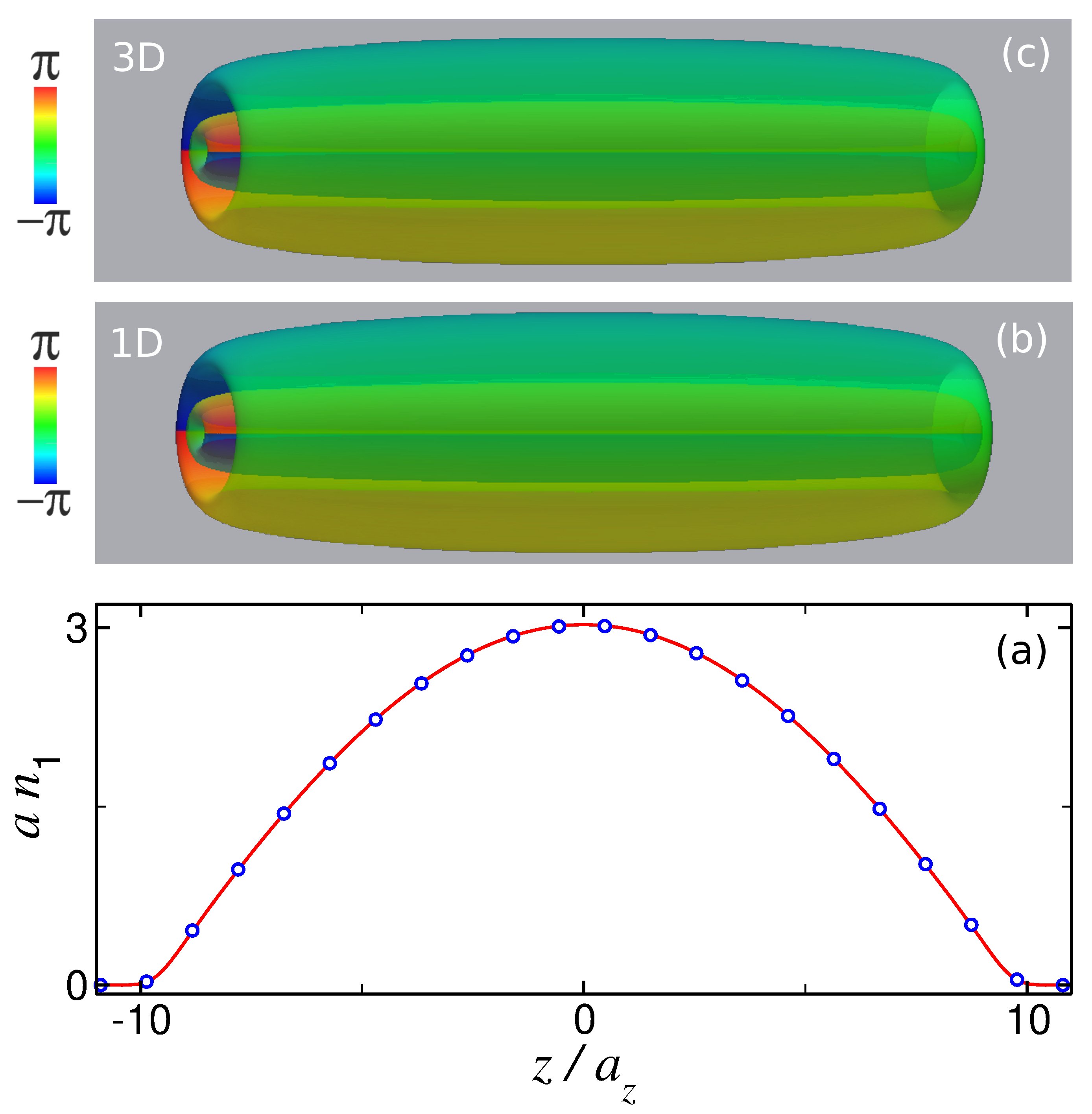}
\end{center}
\caption{
(Color online) Same as Fig. 1 but for a stationary state of the GPE with 
$2\times 10^{4}$ atoms in a $(1,1)$ transversal state with no axial nodes. 
In this case panels (b) and (c) have been radially magnified by a factor 
of $4$. The color maps in these panels show that in both cases the 
condensate phase varies continuously from $-\pi $ to $+\pi $ around the 
vortex axis while it exhibits a $\pi $ shift across the radial node for 
each azimuthal angle.}%
\label{Fig3}%
\end{figure}

%%%%%%%%%%%%%%%%%%%%%%%%%%%%%%%%%%%%%%%%%%%%%%%%%%%%%%%%%%%%%%%%%%%%%%%%%%%%%%%

Figure \ref{Fig3} corresponds to a stationary state of the GPE with
$2\times10^{4}$ atoms in a $(1,1)$ transversal state with no axial nodes. In
this case, at every $z$ plane the atom density features a central unit-charge
vortex as well as a concentric zero-density ring, while the phase displays a
$2\pi$ shift around the vortex singularity and a $\pi$ shift across the radial
node (for each azimuthal angle) [see Figs. \ref{Fig3}(b) and \ref{Fig3}(c)]. 
As occurs with the previous cases, it is apparent that the effective 1D model
(\ref{III-2}) provides accurate predictions for both the axial density
$an_{1}$ [Fig. \ref{Fig3}(a)] and the chemical potential $\mu$ (in this case,
it predicts $\mu=5.58\;\hbar\omega_{\bot}$ with an error less than $0.1\%$)
and also gives a good estimate for the 3D wave function [Figs. \ref{Fig3}(b)
and \ref{Fig3}(c)].

In what follows we shall assume the $^{87}$Rb condensates to be loaded into a
1D optical lattice and look for gap solitons in higher-order transversal
states. Specifically, we consider BECs confined in the radial direction by a
harmonic potential of frequency $\omega_{\bot}=2\pi\times240$ Hz and subject
in the axial direction to a periodic potential of the form%
\begin{equation}
V_{z}(z)=sE_{R}\sin^{2}\left(  \pi z/d\right)  , \label{III-5}%
\end{equation}
with $d=1.55$ $\mu$m being the lattice period and $s=15$ being the lattice
depth in units of the recoil energy $E_{R}\equiv\hbar^{2}\pi^{2}/2Md^{2}$
[which, in turn, sets the typical energy scale of the underlying
(noninteracting) linear problem]. These parameters correspond to a
weak-radial-confinement regime characterized by a recoil energy $E_{R}$ of the
same order as the quantum $\hbar\omega_{\bot}$ ($E_{R}/\hbar\omega_{\bot}=1$).
In such a regime, the energy of gap solitons is always sufficiently large to
excite higher modes of the radial confinement potential and thus 3D
contributions are always relevant. 3D matter-wave gap solitons supported by 3D
optical lattices have been studied in Ref. \cite{Kiv3D}. Here, however, we are
interested in 3D gap solitons supported by 1D lattices, a topic that has
remained largely unexplored \cite{GS-3D}. While matter-wave gap solitons in 1D
lattices have received considerable attention, most studies have focused on an
essentially 1D regime \cite{Kiv3}. However, the 3D weak-radial-confinement
regime is of particular interest for a number of reasons \cite{GS-3D}: (i) Gap
solitons in this regime exhibit a rich radial structure reminiscent of that of
the underlying linear problem. (ii) As a consequence of rotational symmetry,
they can exist even inside the linear energy bands (embedded solitons
\cite{Yang1}). (iii) As occurs in the quasi-1D regime \cite{Wu1-2}, associated
with each 3D linear spectral band there exists a family of \emph{fundamental}
gap solitons (stationary states highly localized in a single lattice site)
that share common radial topological properties with the Bloch waves of the
corresponding linear band; and finally, (iv) contrary to what would be
commonly expected, the weak-radial-confinement regime supports long-lived gap 
solitons.

As is well known, in the noninteracting limit ($N\rightarrow0$), the
stationary states of a BEC in the presence of a 1D lattice are (delocalized)
Bloch waves with energies lying in the allowed bands of the lattice band-gap
spectrum. Because of the separability of axial and radial contributions in the
linear Hamiltonian, the 3D spectrum follows from the 1D band-gap spectrum of
the corresponding axial problem by simply adding the different allowed
energies $E_{m,n_{r}}$ of the radial harmonic oscillator, Eq. (\ref{S-4b}). As
a result, the 3D band-gap spectrum exhibits a series of replicas of the
different 1D energy bands, shifted up in energy by integer multiples of
$\hbar\omega_{\bot}$, with the different shifted bands reflecting the
contribution of the different excited radial modes \cite{GS-3D}. Different
energy bands are characterized by the quantum numbers $(n,m,n_{r})$ where
$n=1,2,3,\ldots$ is the band index of the corresponding 1D axial problem and
$(m,n_{r})$ characterizes the radial state. Based on this fact, we introduced
in Ref. \cite{GS-3D} a nomenclature for gap solitons in the
weak-radial-confinement regime. Indeed, adiabatic continuation of the chemical
potential (or equivalently, the number of atoms $N$) permits connecting the
different states of a given family and thus defines a trajectory in the 
$\mu$-$N$ plane that approaches a certain linear energy band as $N\rightarrow0$. 
This enables us to categorize fundamental gap solitons into families and 
associate a given family with a linear band $(n,m,n_{r})$. Since gap solitons 
of type $(n,m,n_{r})$ can only appear for chemical potentials higher than that 
of the corresponding spectral band, it is clear that there exists a threshold
chemical potential for each family. As $N$ increases or decreases, gap
solitons in a family increase or decrease in size sharing, however, common
topological properties.

%%%%%%%%%%%%%%%%%%%%%%%%%%%%%%%%%%%%%%%%%%%%%%%%%%%%%%%%%%%%%%%%%%%%%%%%%%%%%%%

\begin{figure}[ptb]
\begin{center}
\includegraphics[
width=8.2cm
]{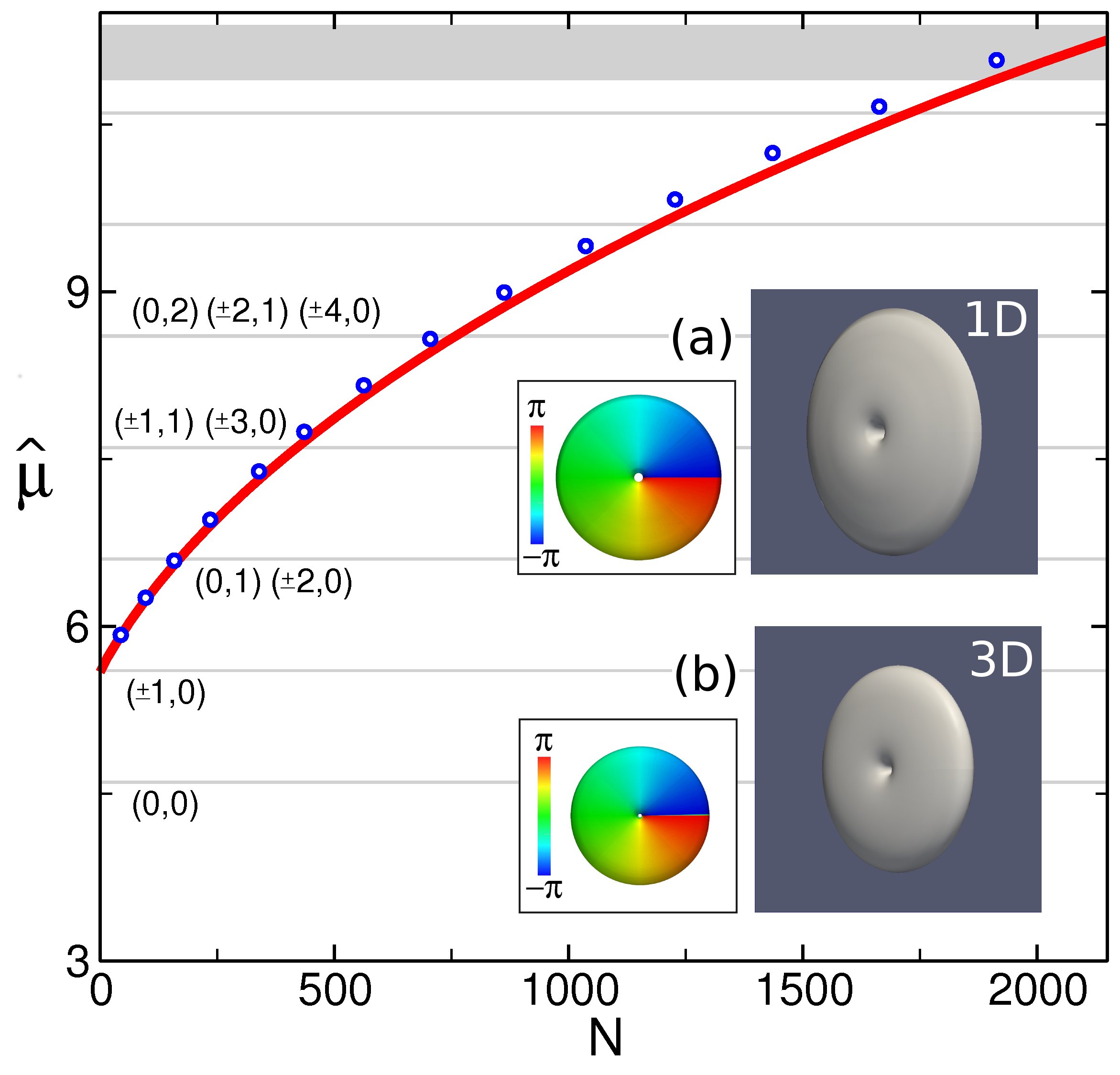}
\end{center}
\caption{
(Color online) Distinctive $\widehat{\mu }(N)$ curve characterizing the
family of $(1,1,0)$ gap solitons, where $\widehat{\mu }$ stands for $\mu
/\hbar \omega _{\bot }$. The solid red line is the result obtained from the
effective 1D model (\ref{III-2}), while open circles have been obtained from
the numerical solution of the full 3D GPE. Horizontal gray strips represent
the spectral bands of the underlying linear problem. The insets display an
example of a soliton of this family with $N=500$ particles. The soliton
wave function as obtained from the effective 1D model (\ref{III-2}) is shown
in panels (a) in terms of an isosurface of the atom density taken at $5\%$
of its maximum (right panel) and a color map of the local phase (left
panel). Panels (b) show the corresponding results as obtained from the 3D
GPE. The field of view in each panel is $6.1$ $\mu $m $\times $ $6.1$ 
$\mu$m. The color maps in panels (a) and (b) show that in both cases the
condensate phase varies continuously from $-\pi $ to $+\pi $ around the
vortex singularity.}%
\label{Fig4}%
\end{figure}

%%%%%%%%%%%%%%%%%%%%%%%%%%%%%%%%%%%%%%%%%%%%%%%%%%%%%%%%%%%%%%%%%%%%%%%%%%%%%%%

Figure \ref{Fig4} shows the trajectory in the $\widehat{\mu}$-$N$ plane
characterizing the family of $(1,1,0)$ gap solitons, where $\widehat{\mu}$
stands for the dimensionless chemical potential $\mu/\hbar\omega_{\bot}$. The
solid red line is the result obtained from the effective 1D model
(\ref{III-2}) while open circles have been obtained from the numerical
solution of the full 3D GPE. Both results agree to within $1.5\%$. In this and
the following figures the horizontal gray strips represent the spectral bands
of the underlying linear problem. The first seven (narrow) strips are
different 3D replicas corresponding to the axial band index $n=1$ [only the
quantum numbers $(m,n_{r})$ are represented in the figures]. The uppermost
(wide) strip is the $(2,0,0)$ spectral band. In the insets we present an
example of a soliton of this family with $N=500$ particles. The soliton
wave function as obtained from the effective 1D model (\ref{III-2}) is shown in
panels (a) in terms of an isosurface of the atom density taken at $5\%$ of its
maximum (right panel) and a color map of\ the local phase (left panel). Panels
(b) show the corresponding results as obtained from the 3D GPE. The field of
view in each panel is $6.1$ $\mu$m $\times$ $6.1$ $\mu$m. Gap solitons in this
family are stationary disk-shaped condensates located at a single lattice site
and featuring a central vortex of unit charge. While condensates with a
similar look have been extensively studied in systems confined in the axial
direction by a strong harmonic potential, it is important to note that gap
solitons are completely different in nature. They are self-trapped nonlinear
waves (bright solitons) resulting from an interplay between nonlinearity and
periodicity which bifurcate from the linear Bloch bands and thus reside in the
band gaps of the lattice spectrum. In particular, $(1,1,0)$ gap solitons are
stationary states of the GPE featuring a vortex of unit charge in the
transversal direction in addition to a (bright) gap soliton in the axial 
direction.

%%%%%%%%%%%%%%%%%%%%%%%%%%%%%%%%%%%%%%%%%%%%%%%%%%%%%%%%%%%%%%%%%%%%%%%%%%%%%%%

\begin{figure}[ptb]
\begin{center}
\includegraphics[
width=8.2cm
]{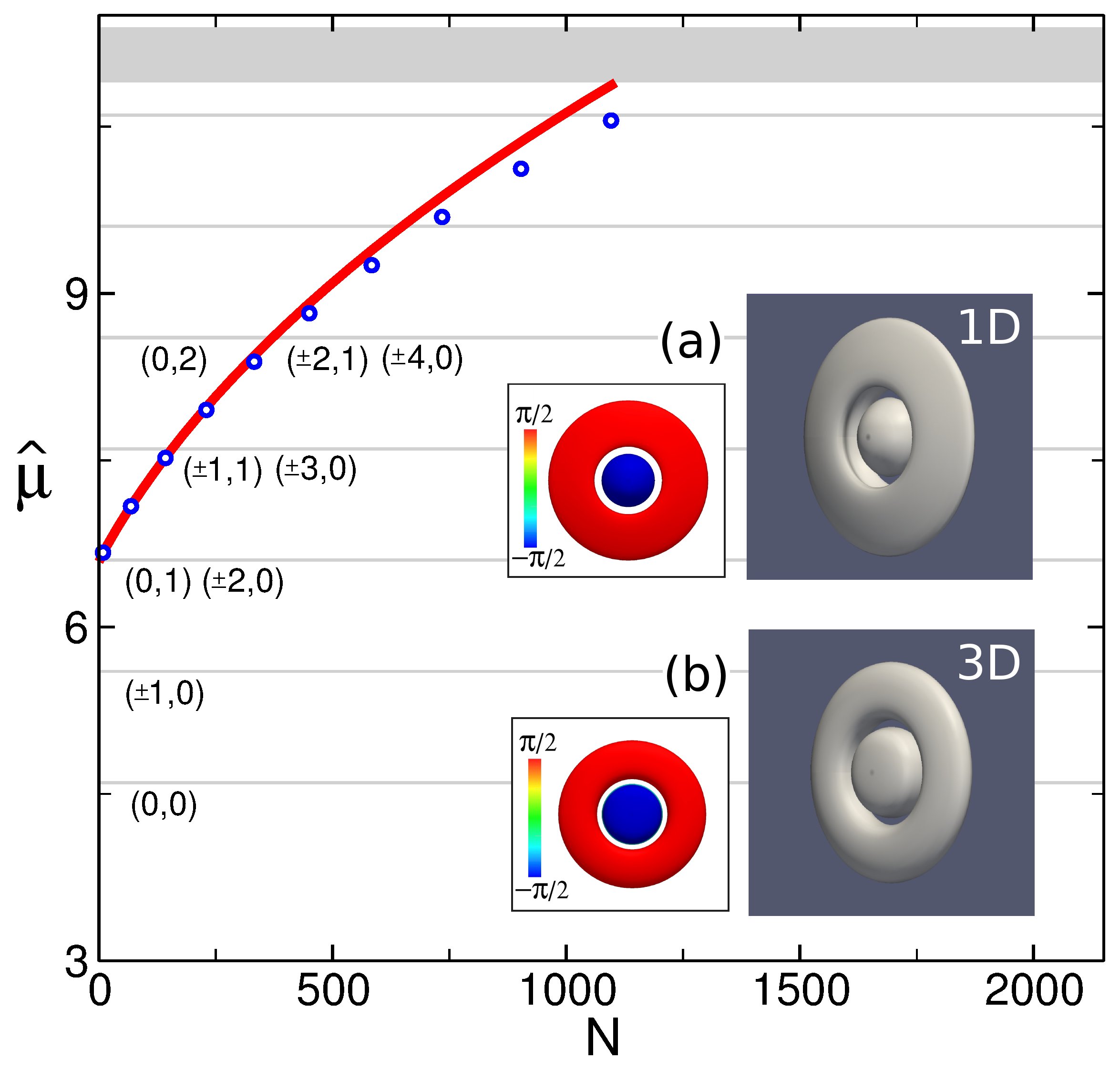}
\end{center}
\caption{
(Color online) Same as Fig. 4 but for the family of $(1,0,1)$ gap solitons.
The color maps in panels (a) and (b) show that in both cases the condensate
phase takes the constant value $-\pi /2$ inside the zero-density ring and 
$+\pi /2$ outside.}%
\label{Fig5}%
\end{figure}

%%%%%%%%%%%%%%%%%%%%%%%%%%%%%%%%%%%%%%%%%%%%%%%%%%%%%%%%%%%%%%%%%%%%%%%%%%%%%%%

Figure \ref{Fig5} displays the distinctive $\widehat{\mu}(N)$ curve of
the$\ (1,0,1)$ family. Even though in this case the prediction from the
effective 1D model (\ref{III-2}) (solid red line) is somewhat less accurate
than before, it still agrees with the 3D results (open circles) to within
$3\%$. As can be seen in the insets, the transversal profile of gap solitons
in this family exhibit a concentric ring of zero density associated with a
$\pi$-phase shift along the radial direction.

%%%%%%%%%%%%%%%%%%%%%%%%%%%%%%%%%%%%%%%%%%%%%%%%%%%%%%%%%%%%%%%%%%%%%%%%%%%%%%%

\begin{figure}[ptb]
\begin{center}
\includegraphics[
width=8.2cm
]{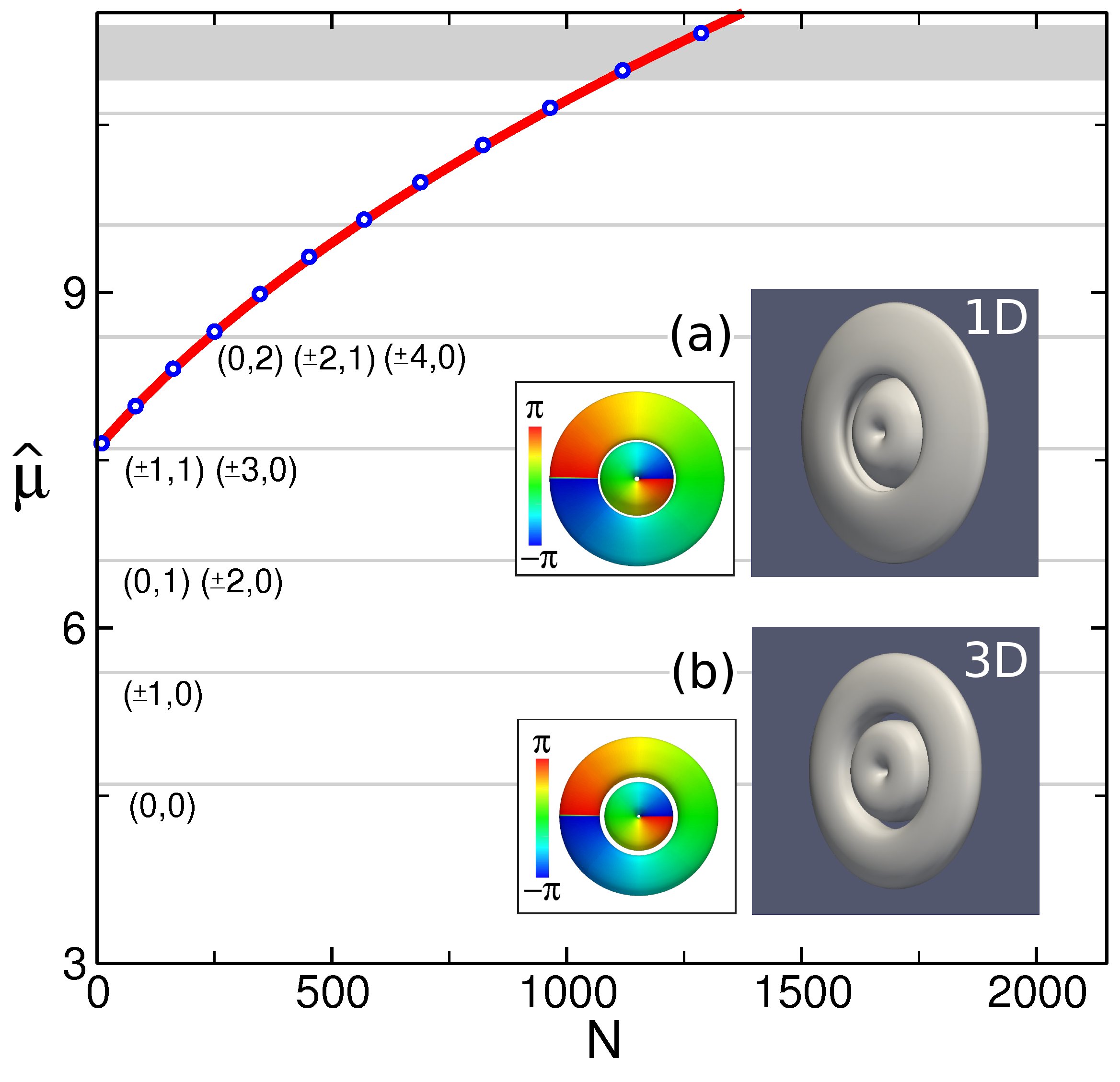}
\end{center}
\caption{
(Color online) Same as Fig. 4 but for the family of $(1,1,1)$ gap solitons.
The color maps in panels (a) and (b) show that in both cases the condensate
phase varies continuously from $-\pi $ to $+\pi $ around the vortex
singularity, while it exhibits a $\pi $ shift across the concentric ring of
zero density.}%
\label{Fig6}%
\end{figure}

%%%%%%%%%%%%%%%%%%%%%%%%%%%%%%%%%%%%%%%%%%%%%%%%%%%%%%%%%%%%%%%%%%%%%%%%%%%%%%%

\begin{figure}[ptb]
\begin{center}
\includegraphics[
width=8.2cm
]{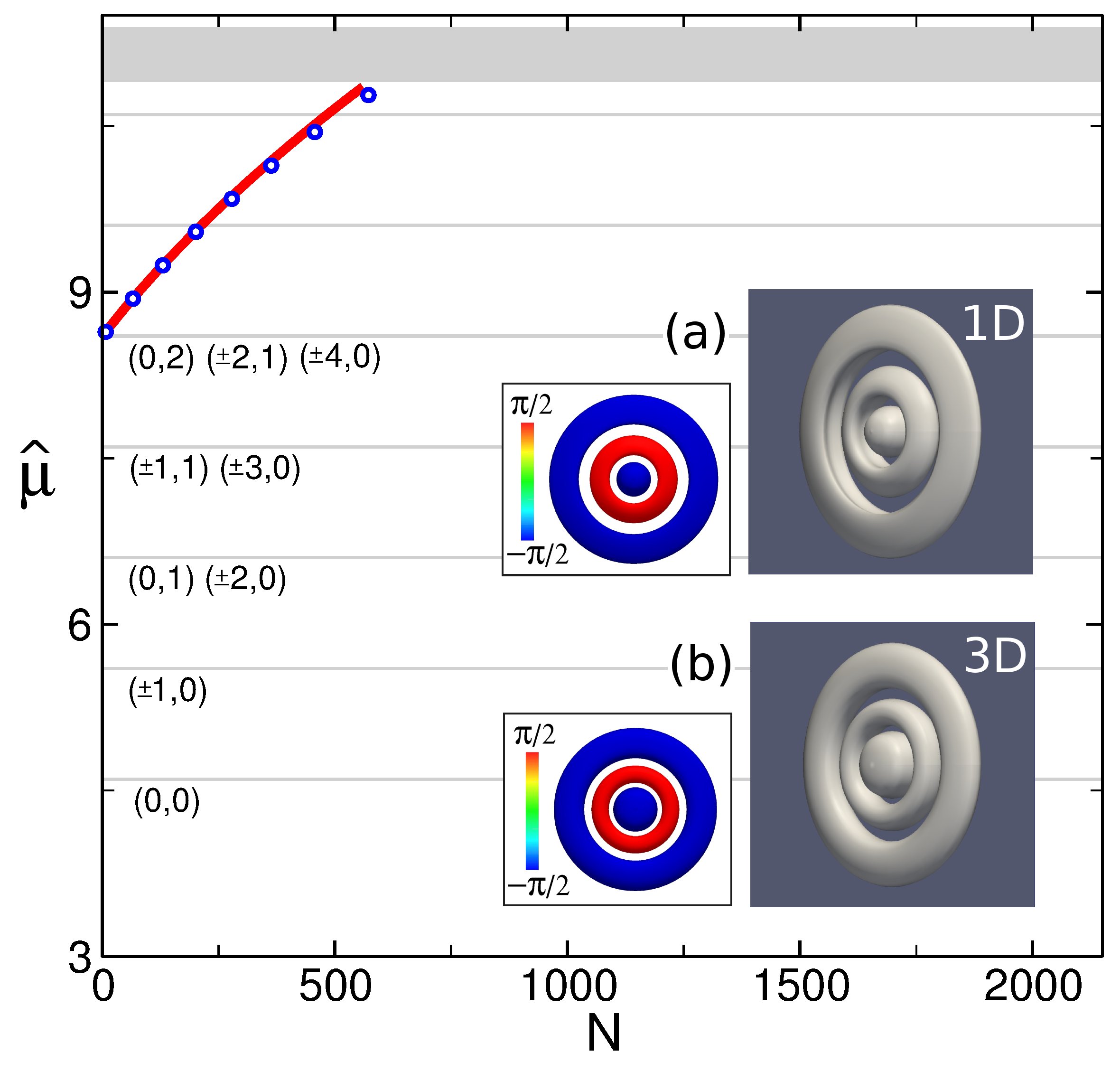}
\end{center}
\caption{
(Color online) Same as Fig. 4 but for the family of $(1,0,2)$ gap solitons.
The color maps in panels (a) and (b) show that in both cases the condensate
phase takes the constant value $-\pi /2$ in the innermost and outermost
regions and $+\pi /2$ in between.}%
\label{Fig7}%
\end{figure}

Finally, predictions for the $(1,1,1)$ and $(1,0,2)$ fundamental families are
presented in Figs. \ref{Fig6} and \ref{Fig7}, respectively. Gap solitons of
type $(1,1,1)$ are self-trapped stationary states of the GPE featuring a
central unit-charge vortex in addition to a concentric zero-density ring,
while those of the $(1,0,2)$ family exhibit two concentric rings of zero
density. As the figures reflect, the threshold chemical potentials for these
families are greater than the previous ones. It is also apparent that the
predicted $\widehat{\mu}(N)$ curves (solid red lines) are in excellent
agreement with the exact numerical 3D results (open circles), being the error
involved less than $1\%$ in both cases. Figures \ref{Fig4}--\ref{Fig7} also
show that the effective 1D model (\ref{III-2}) is able to provide a rather
good estimate for the corresponding 3D wave functions.

\section{IV. SUMMARY AND CONCLUSIONS}

In this work we have demonstrated that an important class of nonlinear
stationary solutions of the 3D GPE in the weak-radial-confinement regime can
be found and characterized in terms of an effective 1D model. This model
provides us with a good approximate solution of the original 3D problem
involving, however, a technical complexity and computational effort similar to
that corresponding to the standard 1D GPE.

%%%%%%%%%%%%%%%%%%%%%%%%%%%%%%%%%%%%%%%%%%%%%%%%%%%%%%%%%%%%%%%%%%%%%%%%%%%%%%%

%%%%%%%%%%%%%%%%%%%%%%%%%%%%%%%%%%%%%%%%%%%%%%%%%%%%%%%%%%%%%%%%%%%%%%%%%%%%%%%

In the weak-radial-confinement regime the system energy is always large enough
so that the 3D GPE admits stationary solutions exhibiting nontrivial
higher-order transversal configurations. In order to address this problem in
terms of an equivalent problem of lower dimensionality we have followed a
variational approach based on both the energy and the chemical-potential
functionals and have derived effective 1D equations that govern the axial
dynamics of BECs in $(m,n_{r})$ transversal states, with $m$ and $n_{r}$
being, respectively, the axial angular momentum and the radial quantum
numbers. Such states feature a central vortex of charge $m$ as well as $n_{r}$
concentric zero-density rings at every $z$ plane.

Interestingly, even though the equations derived in this work generalize
previous proposals to the more complicated case of condensates exhibiting
nontrivial 3D configurations, their functional forms are universal, which is a
consequence of the fact that the specifics of the transversal dynamics can be
absorbed in the renormalization of a couple of parameters ($\alpha$ and
$\gamma$). This permits the study of systems of increasing complexity with the
same computational effort.

The model proposed finds its principal application in those situations where
one is primarily concerned with the existence and classification of axially
localized nonlinear stationary states in a weak-radial-confinement regime. In
these cases the numerical solution of the effective 1D problem provides an
accurate prediction for the axial density profile and the chemical potential
of the 3D condensate and gives a good estimate for the corresponding 3D
wave function as well. In particular, the model proves to be a valuable tool in
the study of the existence and classification of 3D gap solitons supported by
1D optical lattices, where it is able to make very good predictions for the
important $\widehat{\mu}(N)$ curves, which are essential for characterizing
the different fundamental families.

Matter-wave gap solitons in higher-order transversal states are of particular
interest because they are a source for embedded gap solitons and, more
importantly, they either are stable or are expected to be easily stabilized.
In fact, it has been shown that $(1,0,0)$ gap solitons are stable even in the
weak-radial-confinement regime \cite{GS-3D}. The same occurs with the family
$(1,1,0)$ \cite{GS-3D}. Gap solitons of type $(1,m,0)$ with $m>1$ can be
stabilized by simply changing their particle content \cite{GS-3D}. On the
other hand, it has been shown that BECs with a transversal profile similar to
that of the family $(1,0,1)$, confined in a pancake trap, can be stabilized by
adding a second component \cite{Kev-RDS2}. Since the instability of the
$(1,0,1)$ family is purely transversal in nature, the same is expected to
occur in the case of gap solitons. This strategy could also be used to
stabilize gap solitons of type $(1,m,n_{r})$. Alternatively, specifically
engineered pinning potentials might also be used to this purpose \cite{Bosh1}.

While a detailed stability analysis of these nonlinear structures is of great
interest, it requires a separate study with specific computational techniques,
distinct from those required for their generation and characterization. Such
an analysis, which is beyond the scope of this work, will be the subject of a
future publication.

%\begin{acknowledgments}
%This work has been supported by Ministerio de Ciencia e Innovaci\'{o}n 
%(Spain).
%\end{acknowledgments}


\begin{thebibliography}{99}                                                                                               %


\bibitem {Carre2}\textit{Emergent Nonlinear Phenomena in Bose-Einstein
Condensates: Theory and Experiment}, edited by P. G. Kevrekidis, D. J.
Frantzeskakis, and R. Carretero-Gonz\'{a}lez (Springer, Berlin, 2008).

\bibitem {RevDS}D. J. Frantzeskakis, J. Phys. A: Math. Theor. \textbf{43},
213001 (2010).

%=============================================================================================


\bibitem {KivGOP}Y. S. Kivshar and G. P. Agrawal, \textit{Optical 
Solitons: From Fibers to Photonic Crystals} (Academic, San Diego, 2003).

\bibitem {Kiv2D}E. A. Ostrovskaya and Y. S. Kivshar, Phys. Rev. Lett.
\textbf{90}, 160407 (2003).

\bibitem {RMP-Malomed}Y. V. Kartashov, B. A. Malomed, and L. Torner, Rev. Mod.
Phys. \textbf{83}, 247 (2011).

%=============================================================================================


\bibitem {Kon1}V. A. Brazhnyi and V. V. Konotop, Mod. Phys. Lett. B
\textbf{18}, 627 (2004).

\bibitem {Mor1}O. Morsch and M. Oberthaler, Rev. Mod. Phys. \textbf{78}, 179 (2006).

\bibitem {Ober1}B. Eiermann, Th. Anker, M. Albiez, M. Taglieber, P. Treutlein,
K.-P.Marzlin, and M. K. Oberthaler, Phys. Rev. Lett. \textbf{92}, 230401 (2004).

\bibitem {GS1D}F. Kh. Abdullaev, B. B. Baizakov, S. A. Darmanyan, V. V. Konotop and 
M. Salerno, Phys. Rev. A \textbf{64}, 043606 (2001).

\bibitem {Kiv1}P. J. Y. Louis, E. A. Ostrovskaya, C. M. Savage, and Y. S.
Kivshar, Phys. Rev. A \textbf{67}, 013602 (2003).

\bibitem {Abd1}F. Kh. Abdullaev and M. Salerno, Phys. Rev. A \textbf{72},
033617 (2005).

\bibitem {Mal1}T. Mayteevarunyoo and B. A. Malomed, Phys. Rev. A \textbf{74},
033616 (2006).

\bibitem {Wu1-2}Y. Zhang and B. Wu, Phys. Rev. Lett. \textbf{102}, 093905
(2009); Y. Zhang, Z. Liang, and B. Wu, Phys. Rev. A \textbf{80}, 063815 (2009).

\bibitem {VDB7}A. Mu\~{n}oz Mateo, V. Delgado, and B. A. Malomed, Phys. Rev. A
\textbf{83}, 053610 (2011).

\bibitem {Carre1}R. Carretero-Gonz\'{a}lez, D. J. Frantzeskakis, and P. G.
Kevrekidis, Nonlinearity \textbf{21}, R139 (2008).

\bibitem {GS-3D}A. Mu\~{n}oz Mateo, V. Delgado, and B. A. Malomed, Phys. Rev.
A \textbf{82}, 053606 (2010).

%=============================================================================================


%\cite{Olsha1,Kett1,Das1,Strin1,VDB1,Guerin1,Isa1}


\bibitem {Olsha1}M. Olshanii, Phys. Rev. Lett. \textbf{81}, 938 (1998); D. S.
Petrov, G. V. Shlyapnikov, and J. T. M. Walraven, \textit{ibid.} \textbf{85},
3745 (2000); V. Dunjko, V. Lorent, and M. Olshanii, \textit{ibid.}
\textbf{86}, 5413 (2001).

\bibitem {Kett1}A. G\"{o}rlitz, J. M. Vogels, A. E. Leanhardt, C. Raman, T. L.
Gustavson, J. R. Abo-Shaeer, A. P. Chikkatur, S. Gupta, S. Inouye, T.
Rosenband, and W. Ketterle, Phys. Rev. Lett. \textbf{87}, 130402 (2001).

\bibitem {Das1}K. K. Das, Phys. Rev. A \textbf{66}, 053612 (2002).

\bibitem {Strin1}C. Menotti and S. Stringari, Phys. Rev. A \textbf{66}, 043610
(2002); J. N. Fuchs, X. Leyronas, and R. Combescot, \textit{ibid.}
\textbf{68}, 043610 (2003); F. Gerbier, Europhys. Lett. \textbf{66}, 771 (2004).

\bibitem {VDB1}A. Mu\~{n}oz Mateo and V. Delgado, Phys. Rev. A \textbf{74},
065602 (2006).

\bibitem {Guerin1}W. Guerin, J.-F. Riou, J. P. Gaebler, V. Josse, P. Bouyer,
and A. Aspect, Phys. Rev. Lett. \textbf{97}, 200402 (2006).

\bibitem {Isa1}J. Armijo, T. Jacqmin, K. Kheruntsyan, and I. Bouchoule, Phys.
Rev. A \textbf{83}, 021605(R) (2011); J. Armijo, Phys. Rev. Lett.
\textbf{108}, 225306 (2012).

%=============================================================================================


\bibitem {Shin2}Y. Shin, M. Saba, T. A. Pasquini, W. Ketterle, D. E.
Pritchard, and A. E. Leanhardt, Phys. Rev. Lett. \textbf{92}, 050405 (2004).

\bibitem {Schumm1}T. Schumm, S. Hofferberth, L. M. Andersson, S. Wildermuth,
S. Groth, I. Bar-Joseph, J. Schmiedmayer and P. Kr\"{u}ger, Nat. Phys.
\textbf{1}, 57 (2005).

\bibitem {Wang1}Y.-J. Wang, D. Z. Anderson, V. M. Bright, E. A. Cornell, Q.
Diot, T. Kishimoto, M. Prentiss, R. A. Saravanan, S. R. Segal, and S. Wu,
Phys. Rev. Lett. \textbf{94}, 090405 (2005).

\bibitem {Burger1}S. Burger, K. Bongs, S. Dettmer, W. Ertmer, K. Sengstock, A.
Sanpera, G. V. Shlyapnikov, and M. Lewenstein, Phys. Rev. Lett. \textbf{83},
5198 (1999).

\bibitem {Brand1}L. D. Carr, J. Brand, S. Burger, and A. Sanpera, Phys. Rev. A
\textbf{63}, 051601(R) (2001).

\bibitem {PRL06}A. Mu\~{n}oz Mateo and V. Delgado, Phys. Rev. Lett.
\textbf{97}, 180409 (2006).

\bibitem {Mott}J. A. M. Huhtam\"{a}ki, M. M\"{o}tt\"{o}nen, T. Isoshima, V.
Pietil\"{a}, and S. M. M. Virtanen, Phys. Rev. Lett. \textbf{97}, 110406 (2006).

%=============================================================================================


%\cite{Vic1,Jack1,Chio1,Gora,Salas1,Kam1,Clark1,Yuka1,Ef1DEqs,Nico1}


\bibitem {Vic1}V. M. P\'{e}rez-Garc\'{\i}a, H. Michinel, J. I. Cirac, M.
Lewenstein, and P. Zoller, Phys. Rev. A \textbf{56}, 1424 (1997).

\bibitem {Jack1}A. D. Jackson, G. M. Kavoulakis, and C. J. Pethick, Phys. Rev.
A \textbf{58}, 2417 (1998).

\bibitem {Chio1}M. L. Chiofalo and M. P. Tosi, Phys. Lett. A \textbf{268}, 406 (2000).

\bibitem {Gora}A. E. Muryshev, G. V. Shlyapnikov, W. Ertmer, K. Sengstock, and
M. Lewenstein, Phys. Rev. Lett. \textbf{89}, 110401 (2002).

\bibitem {Salas1}L. Salasnich, A. Parola, and L. Reatto, Phys. Rev. A
\textbf{65}, 043614 (2002); L. Salasnich, Laser Phys. \textbf{12}, 198 (2002).

\bibitem {Kam1}A. M. Kamchatnov and V. S. Shchesnovich, Phys. Rev. A
\textbf{70}, 023604 (2004).

\bibitem {Clark1}M. Edwards, L. M. DeBeer, M. Demenikov, J. Galbreath, T. J.
Mahaney, B. Nelsen, and C. W. Clark, J. Phys. B: At. Mol. Opt. Phys.
\textbf{38}, 363 (2005).

\bibitem {Yuka1}K.-P. Marzlin and V. I. Yukalov, Eur. Phys. J. D \textbf{33},
253 (2005).

\bibitem {Ef1DEqs}A. Mu\~{n}oz Mateo and V. Delgado, Phys. Rev. A \textbf{75},
063610 (2007); \textbf{77}, 013617 (2008); Ann. Phys. \textbf{324}, 709 (2009).

\bibitem {Nico1}A. I. Nicolin and R. Carretero-Gonz\'{a}lez, Physica A
\textbf{387}, 6032 (2008); A. I. Nicolin and M. C. Raportaru, \textit{ibid}. 
\textbf{389}, 4663 (2010); A. I. Nicolin, \textit{ibid}. \textbf{391}, 1062
(2012); A. Bala\v{z} and A. I. Nicolin, Phys. Rev. A \textbf{85}, 023613 (2012).

%=============================================================================================


%\cite{Modug1,Salas2,Salas3,Salas4,Adhik1,Adhik2,Adhik3,Carre3,Salas5,Adhik4}


\bibitem {Modug1}P. Massignan and M. Modugno, Phys. Rev. A \textbf{67}, 023614
(2003); W. Zhang and L. You, \textit{ibid.} \textbf{71}, 025603 (2005).

\bibitem {Salas2}L. Salasnich and B. A. Malomed, Phys. Rev. A \textbf{74},
053610 (2006); L. Salasnich, A. Cetoli, B. A. Malomed, F. Toigo, and L.
Reatto, \textit{ibid.} \textbf{76}, 013623 (2007).

\bibitem {Salas3}L. Salasnich, B. A. Malomed, and F. Toigo, and  Phys. Rev. A
\textbf{76}, 063614 (2007).

\bibitem {Salas4}L. Salasnich, J. Phys. A: Math. Theor. \textbf{42}, 335205 (2009).

\bibitem {Adhik1}C. A. G. Buitrago and S. K. Adhikari, J. Phys. B: At. Mol.
Opt. Phys. \textbf{42}, 215306 (2009).

\bibitem {Adhik2}Luis E. Young-S., L. Salasnich, and S. K. Adhikari, Phys. Rev.
A \textbf{82}, 053601 (2010).

\bibitem {Adhik3}S. K. Adhikari, J. Phys. B: At. Mol. Opt. Phys. \textbf{44},
075301 (2011).

\bibitem {Carre3}S. Middelkamp, J. J. Chang, C. Hamner, R.
Carretero-Gonz\'{a}lez, P. G. Kevrekidis, V. Achilleos, D. J. Frantzeskakis,
P. Schmelcher, and P. Engels, Phys. Lett. A \textbf{375}, 642 (2011).

\bibitem {Salas5}L. Salasnich and B. A. Malomed, J. Phys. B: At. Mol. Opt.
Phys. \textbf{45}, 055302, (2012).

\bibitem {Adhik4}P. Muruganandam, and S. K. Adhikari, Laser Phys. \textbf{22},
813, (2012).

%=============================================================================================


%Experimental confirmation
%\cite{Carre3,Kev1,Wel1,Kev2}


\bibitem {Kev1}S. Middelkamp, G. Theocharis, P. G. Kevrekidis, D. J.
Frantzeskakis, and P. Schmelcher, Phys. Rev. A \textbf{81}, 053618 (2010).

\bibitem {Wel1}G. Theocharis, A. Weller, J. P. Ronzheimer, C. Gross, M. K.
Oberthaler, P. G. Kevrekidis, and D. J. Frantzeskakis, Phys. Rev. A
\textbf{81}, 063604 (2010).

\bibitem {Kev2}C. Wang, P. G. Kevrekidis, T. P. Horikis, D. J. Frantzeskakis,
Phys. Lett. A, \textbf{374}, 3863 (2010).

%=============================================================================================


%\cite{Kiv-RDS,LCarr-RDS,Kev-RDS1,San-RDS,Kev-RDS2,Liu-RDS}


\bibitem {Kiv-RDS}G. Theocharis, D. J. Frantzeskakis, P.G. Kevrekidis, B. A.
Malomed, and Y. S. Kivshar, Phys. Rev. Lett. \textbf{90}, 120403 (2003).

\bibitem {LCarr-RDS}L. D. Carr and C. W. Clark, Phys. Rev. Lett. \textbf{97},
010403 (2006); Phys. Rev. A \textbf{74}, 043613 (2006).

\bibitem {Kev-RDS1}K. M. Mertes, J.W. Merrill, R. Carretero-Gonz\'{a}lez, D.
J. Frantzeskakis, P. G. Kevrekidis, and D. S. Hall, Phys. Rev. Lett.
\textbf{99}, 190402 (2007).

\bibitem {San-RDS}M. Scherer, B. L\"{u}cke, G. Gebreyesus, O. Topic, F.
Deuretzbacher, W. Ertmer, L. Santos, J. J. Arlt, and C. Klempt, Phys. Rev.
Lett. \textbf{105}, 135302 (2010).

\bibitem {Kev-RDS2}J. Stockhofe, P. G. Kevrekidis, D. J. Frantzeskakis, P.
Schmelcher, J. Phys. B: At. Mol. Opt. Phys. \textbf{44}, 191003 (2011).

\bibitem {Liu-RDS}J. Li, D. S. Wang, Z. Y. Wu, Y. M. Yu, and W. M. Liu, Phys.
Rev. A \textbf{86}, 023628 (2012).

%=============================================================================================


\bibitem {GPE}E. P. Gross, Nuovo Cimento \textbf{20}, 454 (1961); J. Math.
Phys. \textbf{4}, 195 (1963); L. P. Pitaevskii, Zh. Eksp. Teor. Fiz.
\textbf{40}, 646 (1961) [Sov. Phys. JETP \textbf{13}, 451 (1961)].

\bibitem {Abramo}M. Abramowitz and I. Stegun, \emph{Handbook of Mathematical
Functions} (Dover, New York, 1972).

\bibitem {Kiv3D}T. J. Alexander, E. A. Ostrovskaya, A. A. Sukhorukov, and Y.
S. Kivshar, Phys. Rev. A \textbf{72}, 043603 (2005).

\bibitem {Kiv3}M. Matuszewski, W. Kr\'{o}likowski, M. Trippenbach, and Y. S.
Kivshar, Phys. Rev. A \textbf{73}, 063621 (2006).

%=============================================================================================


\bibitem {Yang1}J. Yang, B. A. Malomed, and D. J. Kaup, Phys. Rev. Lett.
\textbf{83}, 1958 (1999); A. R. Champneys, B. A. Malomed, J. Yang, and D. J.
Kaup, Physica D \textbf{152-153}, 340 (2001); J. Yang, Phys. Rev. A
\textbf{82}, 053828 (2010).

%=============================================================================================


\bibitem {Bosh1}K. Henderson, C. Ryu, C. MacCormick, and M. G. Boshier, New J.
Phys. \textbf{11}, 043030 (2009).
\end{thebibliography}
\end{document}